\documentclass[preprint,showpacs,preprintnumbers,amsmath,amssymb]{revtex4}
\usepackage{graphicx}
\usepackage{dcolumn}
\usepackage{bm}
\input epsf

\begin{document}



\title{A dark radiation era prior to the inflationary phase}

\author{Sergio del Campo$^{1}$}
\email{sdelcamp@ucv.cl}

\author{Ram\'on Herrera$^{1}$}
\email{ramon.herrera@ucv.cl}

\author{V\'{\i}ctor H. C\'{a}rdenas$^{2}$}
\email{victor@dfa.uv.cl}

\affiliation{$^{1}$Instituto de F\'{\i}sica, Pontificia
Universidad Cat\'olica de Valpara\'{\i}so, Casilla 4059,
Valpara\'{\i}so, Chile}

\affiliation{$^2$ Departamento de F\'{\i}sica y Astronomia,
Universidad de Valpara\'{\i}so, Gran Bretana 1111,
Valpara\'{\i}so, Chile}

\begin{abstract}
A cosmological model dominated at the beginning by a dark
radiation followed by a period of inflation is presented. This
model is based on a Randall-Sundrum II type brane-world. Current
observational data are used to fix the parameters associated to
the dark radiation.
\end{abstract}


\maketitle

\section{Introduction}

One interesting question to ask is: What was the previous period
to the inflationary phase? or similarly, What was the initial
condition that make inflation to occur? The obscureness of the
physical mechanism underlying inflation results quite frustrating.
Nowadays, the beginning and  end of the inflationary period
result, by themselves, theoretical challenges by themselves. In
this respect, some researchers have tried to give answer to the
above question. For instance, there have been attempts to find an
initial state for the inflationary universe models from the
quantum cosmology. Here, the so-called wave function of the
universe \cite{WF} has been advised for doing this task.
Furthermore, there have been attempts to address this question
through the introduction of the "emerging universe". An emergent
universe is a model universe in which there is no timelike
singularity, having almost static behavior in the infinite past
($t\rightarrow -\infty$) \cite{E}. It is assumed that the initial
conditions are specified so as the static configuration represents
the past eternal state of the universe, out of which the universe
slowly evolves into an inflationary phase. Different schemes  have
been used for describing this sort of models. In Ref.\cite{BW} a
brane world scenario was considered. Moreover, a
Jordan-Brans-Dicke sort of theory has been also considered
\cite{JBD}. On the other hand, the pre-big-bang cosmology inspired
by  the superstring theories has been suggested as a possible
implementation of the inflationary universe scenario \cite{PBB}.
However, the scenario validity  as a viable inflationary model has
been questioned on the grounds of its initial conditions \cite{L}.
We have not firm conclusions about this point at the moment.

On the other hand, the cyclic/oscillatory universes have been
present in cosmology for a long time \cite{T}. Here, the main
attraction was that the initial conditions could in principle be
avoided. However, these models show severe construction
difficulties  within the  general relativity context. First of
all, the number of bounces in the past are restricted to entropy
constraints; nevertheless,  the main difficulty is perhaps that
any bounce would be singular, thus resulting in the breakdown of
the general relativity. Recent developments on brane world
scenarios have renewed interest on the  cyclic/oscillatory
universes \cite{O}.  The cyclic/oscillatory universe that
ultimately undergoes inflationary expansion after a finite number
of cycles has also been investigated \cite{3}. However, a physical
mechanism is missing for inducing the corresponding number of
bounces. Some authors have considered the loop quantum cosmology
in order to solve this problem \cite{LQC}. Nevertheless, the
oscillations in such models can be characterized  by reformulating
the semi–classical dynamics in terms of an effective phantom
fluid.

The idea that our universe is confined to a brane in a higher
dimensional bulk spacetime has been constantly under study  in the
last years \cite{branecos}. One of the most studied model is the
so-called Randall-Sundrum \cite{RS} (usually cited as RSII model),
in which our universe lived in a brane embedded in an Anti-de
Sitter (AdS) five-dimensional bulk spacetime. Soon after its
appearance, a lot of work was made to built cosmological
extensions of the RS II model \cite{CosB}. One of the most
relevant consequences of these studies was the modification of the
Friedmann equation for energy densities of the order of the brane
tension,  and also  the appearance of an additional term, usually
called dark radiation, in addition \cite{M}. This latter term,
i.e. the dark radiation, results crucial, as its presence can
spoil the big bang nucleosynthesis \cite{N} or even modify  the
overall amplitude of the fluctuations spectrum   associated to
large scale structure formation processes.  Thus, it seems natural
to look for realizations of inflation where the dark radiation
plays a role.

In this Letter,  a model of the early universe is proposed, where
the dark radiation is taken into account. In this context,  the
effects that the  dark radiation might have previously to the
inflationary period would be studied. Actually, this transition
will be possible for acceptable values  of the  parameters that
appear in the model.

The present Letter is organized as follows. The basic field
equations are formulated in Section II. Section III involves the
dark radiation dominated era.  The inflationary phase is described
in Section IV , where some astronomical data are used in order to
fix some parameters. Finally,  the results of the Letter are
summarized in Section V.

\section{The Basic Equations}
The field equations are recasts in the following convenient form
through a flat Friedmann-Robertson-Walker metric (FRW), in which
the brane-bulk energy exchange is included (for detail see
Ref.\cite{KKTTZ})
\begin{equation}\label{1}
\dot{\rho} + 3H(\rho + p) = -T(\rho),
\end{equation}
\begin{equation}\label{2}
\dot{\chi} + 4H\chi =   \left(\frac{\rho}{\sigma} + 1
\right)T(\rho),
\end{equation}
and
\begin{equation}\label{3}
H^{2} = 2\gamma \rho \left( 1+ \frac{\rho}{2 \sigma}\right) +
2\gamma\chi + \Lambda ^{(4)}.
\end{equation}
Here, the constants  $\sigma$,  $\Lambda ^{(4)}$ and
$\gamma=4\pi\,G=4\pi/m_p^2$ are related to  the brane tension, the
effective cosmological constant on the brane and the Newton
constant or equivalently, the Planck mass in 4-dimensions,
respectively. On the other hand, $T(\rho)=2T^{0}_{4}$ is the
discontinuity of the 04 component of the bulk energy momentum
tensor and the matter is represented by a perfect fluid which is
characterized by a pressure $p$ and energy density $\rho$, which
are related by a specific equation of state. The energy density
$\chi$ represents the dark or the mirage radiation\cite{KK}, and
$H=\frac{\dot{a}}{a}$ the Hubble factor, with $a=a(t)$ the scale
factor. Dots here represent derivative with respect to the
cosmological time $t$.

It is interesting to note that if we assume some equation of state
parameter $\omega$, defined as $\omega\equiv \frac{p}{\rho}$ and
vanishing cosmological constant, $\Lambda ^{(4)}=0$, from the set
of Eqs. (\ref{1})-(\ref{3}), we get that
\begin{equation}\label{eq1}
\rho^2-\left(\frac{1-3\omega}{1+3\omega}\right)\sigma \rho =
-\left(\frac{1}{1+3\omega}\right)\frac{\sigma}{\gamma}\left(\dot{H}
+2 H^2\right) ,
\end{equation}
which tells us how the matter energy density, $\rho$,
cosmologically evolves
 for an arbitrary bulk energy momentum tensor,
$T(\rho)$, i.e. we could get $\rho$ as an explicit function of the
cosmological time, $t$, after knowing the scale factor, $a(t)$.
Here, we have assumed that $\omega$ is different to $-1/3$.

We could give an intuitive picture about our approach if we
combine Eqs. (\ref{1}) and (\ref{2}) in such a way that we write
down  a single equation
\begin{equation} \label{eq3}
\left[ \dot{\rho}+3 H (\rho + p) \right]\left(
1+\frac{\rho}{\sigma}\right)= - \left( \dot{\chi} + 4 H \chi
\right).
\end{equation}
In this respect, our fundamental set of equations becomes Eqs.
(\ref{3}) and (\ref{eq3}). We see that this set is equivalent to
the set of Eqs. (8) and (9) of Ref.\cite{g2}, if we take the
curvature parameter $k=0$, and we identify $\chi$ with $ \cal{M} $
(the "generalized comoving mass") via $\chi=\frac{12 M^2}{\pi^2
V}\frac{\cal{M}}{a^4}$ and $p_{D}=\frac{1}{3}\chi$ (see Ref.
\cite{g2} for the meaning of the scripts that appear in these
expressions). In this picture $\cal{M}$  gives information about
the bulk-effect corrections on the brane, when the bulk spacetime
contains an arbitrary matter configuration\cite{g1}. In this
sense, following Ref.\cite{g2} our case would correspond to a
configuration in which $p_{D}=\frac{1}{3}\chi$, thus representing
a AdS-Vaidya bulk\cite{AV,g3} with energy exchange between the
brane and a radiation field in the bulk in the high-energy
regime\cite{g4}.

Let us write down the field equations (\ref{1}) and (\ref{2}) in
terms of the scale factor derivatives through the relation
$dt=da/Ha$, represented by primes, so that these Equations take
the form
\begin{equation}
\rho^{\prime}+\frac{3}{a}(1+\omega)\rho=-\frac{T(\rho)}{Ha},\label{roeq2}%
\end{equation}%
and
\begin{equation}
\chi^{\prime}+\frac{4}{a}\chi=\left(\frac{\rho}{\sigma}+1\right)\frac{T(\rho)}{Ha},
\label{chieq2}
\end{equation}
respectively. These two latter equations along with the Friedmann
equation, equation (\ref{3}), form the basic Equations that we
would like to take into account  to describe a facet in which the
beginning of the universe is dominated by the dark radiation
energy. This situation will be processed in the following section.

\section{ General solutions  and Dark radiation dominated era}

In order to find a solution to the previous set of field
equations, let us begin supposing the following ansatz for the
energy density, $\rho$, as a function of the scale factor $a$,
\begin{equation}
\rho(a)=\frac{B}{a^{3(1+\omega)}}+\frac{A}{a^{\alpha}},\label{ansatz}%
\end{equation}
where $A$ and $B$  are two arbitrary constants. Note that the
first term corresponds to the solution of the homogeneous Eq.
(\ref{roeq2}), i.e. when $T(\rho)=0$. Therefore, the second term
could be taken as a solution of the inhomogeneous part of this
equation, with $\alpha$ an arbitrary constant.

Substituting Eq. (\ref{ansatz}) into (\ref{roeq2}) we find for the interaction term%
\begin{equation}
T=\frac{HA}{a^{\alpha}}\left[
\alpha-3(1+\omega)\right],\label{interac}
\end{equation}
with $\alpha \neq 3(1+\omega)$.

Using Eqs. (\ref{ansatz}) and (\ref{interac}) into Eq.
(\ref{chieq2}), we find for the dark radiation density, $\chi$,
the following solution
\begin{equation}
\chi(a)=\frac{\tilde{A}}{a^{2\alpha}}+\frac{\tilde{B}}{a^{\mu-1}}+\frac
{\tilde{C}}{a^{\alpha}}+\frac{D}{a^{4}},\label{chisol}%
\end{equation}
where
\begin{equation}
\tilde{A}  =\frac{A^2}{\sigma (4-2\alpha)}\left[
\alpha-3(1+\omega)\right], \alpha \neq 2,
,\label{ctea}\\
\end{equation}
\begin{equation}\label{cteb}
\tilde{B}  =\frac{A B}{\sigma (5-\mu)}\left[  \alpha-3(1+\omega)\right], \mu \neq 5\\
\end{equation}
\begin{equation}\label{ctec}
\tilde{C}  =\frac{A}{4-\alpha}\left[ \alpha-3(1+\omega )\right],
\alpha \neq 4,
\end{equation}
and $D$ is an integration constant. Here,
$\mu-1=3(1+\omega)+\alpha$. Note that if $T(\rho)=0$, i.e.
$\alpha=3(1+\omega)$, then only the last term of Eq.
(\ref{chisol}) survives, and thus $\chi\sim a^{-4}$, which
corresponds to the solution of the homogeneous equation for the
dark radiation. In the following analysis, we will assume that
$\alpha>3(1+\omega)$. As we will see in the next section, this
assumption will simplify our analysis.

Regarding these solutions, i.e. Eqs. (\ref{ansatz}) and
(\ref{chisol}), let us assume that  the universe is initially in a
high energy state in which the dark radiation component dominating
the universe, $\chi\sim a^{-4}\gg a^{-2\alpha}\sim \rho^2$ where
$\alpha<2$. Note that the energy exchange term $T$ given by
Eq.(\ref{interac}) is related with the spatial discontinuity of
the bulk metric $a(t, \eta)$ (here $\eta$ represents the bulk
coordinate) near the brane and must be positive in order the
embedding of the 3-brane could be possible (see
Refs.\cite{g2,AV}). Therefore, a positive $T$ gives
$\alpha>3(1+\omega)$, and together with the dominant dark
radiation period, $\alpha<2$, from which we get that
$\omega<-1/3$, we could get an inflationary period. If, on the
other hand, $\alpha<3(1+\omega)$ and $\alpha>2$ (which yields to
$\omega>-1/3$)  the universe is consistent only with  a negative
$T$, this  energy transfer from the brane to the bulk cannot lead
to periods of accelerated expansion on the brane \cite{7}. Here
$T\sim a^{-(2+\alpha)}$, so that
\begin{equation}
\chi \gg \sqrt{\gamma}\,T \gg \frac{\rho^2}{2 \sigma }  \gg
\rho.\label{ic}
\end{equation}

At this stage, the last term in Eq. (\ref{chisol}) dominates, and
thus, the Friedmann equation can be written as followed
\begin{equation}
H^2 \simeq 2\gamma \chi,
\end{equation}
and the evolution of the dark radiation density is governed by the
Equation
\begin{equation}
\dot{\chi} + 4H\chi \simeq \frac{\rho}{\sigma}   T(\rho).
\end{equation}

An attractor solution for this configuration is that of a
power-law scale factor $a(t)\sim t^n$, with $n<1$ . In this regime
$\rho \simeq t^{-1}$ and $\chi \simeq t^{-2}$. The interaction
term behaves as $T(t) \simeq t^{-2}$ which gives $T(\rho) \sim
\rho^{2}$. This period is prolonged until a specific time, say
$t_e$, that corresponds to when it is satisfied $\chi(t_e)\approx
\sqrt{\gamma}\,T(t_e)\approx\frac{\rho^2(t_e)}{2 \sigma}\sim
a_e^{-2\alpha}$, where $a_e=a(t_e)$. Thus we have a dark radiation
dominated period, which starts at some initial time, $t_i$, and
finishes at the time $t_e$, i.e. we have $t_i\leq t \leq t_e$ for
the dark radiation period.


We will restrict ourselves to a phenomenological expressions for
the bulk-energy momentum tensor, $T$, previously considered in the
literature from now on. One common choice for this is $T=3\xi H
\rho$, with $\xi$ a positive-definite dimensionless constant,
which is taken to be $\xi\ll 1$. This choice has been studied in
detail in Ref.\cite{yo}. Then, the field equations become the dark
radiation dominated era, i.e. for $\chi\gg \frac{\rho^2}{2
\sigma}$
\begin{equation}\label{16}
H^2 \simeq 2\gamma \chi,
\end{equation}
\begin{equation}\label{17}
\dot{\chi} + 4H\chi \simeq 0,
\end{equation}
and
\begin{equation}\label{11}
\dot{\rho} + 3H\rho(1 + \omega) = -3\xi H \rho.
\end{equation}
The apparent asymmetry between (\ref{17}) and (\ref{11}) needs a
clarification. Assuming the dark radiation domination occurs in a
high energy regime implies that $\chi\gg \rho (1+\rho/2\sigma) $,
as can be see comparing Eq. (\ref{16}) with Eq. (\ref{3}). Using
the interaction as $T=3\xi H \rho$, the same factor appears at the
right hand of Eq. (\ref{17}), being negligible small compared to
the homogeneous term $4H\chi$ in the same Equation. So, the system
(\ref{16},\ref{17},\ref{11}) is a consistently high energy limit
of the original system of equations, with an interaction term
different from zero; i.e. $T(\rho) \neq 0$.

The corresponding solutions of these Equations are given by
\begin{equation}\label{19}
\rho(a)=\widetilde{\rho}
(\frac{\widetilde{a}}{a})^{3(\xi+\omega+1)},
\end{equation}
where tilde specifies some time well inside of the dark radiation
period, i.e. $\widetilde{t}\ll t_e$, and $\chi(a) \approx
\frac{D}{a^{4}}$, with $a(t)\sim t^{1/2}$. Having described the
period once was dominated by radiation, we will proceed to
consider a posterior period of inflation. This will allow fixing
some parameters that define our model in the radiation-dominated
period, using astronomic data.  We should stress  here that the
aim of the present paper is to put some constraint on the dark
radiation term ($\chi\sim D/a^4$) previous to the inflationary
period ($\rho^2$) independently if the bulk-energy momentum tensor
$T$ could be vanished or not. We assume here that $T$ is
negligible in the two epochs, dark radiation and inflationary
periods.

\section{Inflationary Phase}

Following an approach similar to that described in Ref.\cite{vh},
the proper transition from a dark radiation dominated regime to an
accelerated one is through the solution \cite{20}
\begin{equation}
a(t)= a_{in} \left[ \sinh(C_{2}\ln (1+t/t_{in}))\right]^{1/n},
\end{equation}
which \textit{interpolates} between the stage $t /t_{in}\ll 1$,
where the scale factor follows the law
\begin{equation}\label{arad}
a(t) \simeq a_{in}\,(C_{2})^{1/n}(t/t_{in})^{1/n},
\end{equation}
and the power-law inflationary phase, in which  $t/t_{in} \gg 1$,
and the scale factor becomes
\begin{equation}
a(t) \simeq
a_{in}\,(1+t/t_{in})^{C_{2}/n}\simeq\,a_{in}\,\,(t/t_{in})^{C_{2}/n}
=\,a_{in}\,(t/t_{in})^{p}.\label{power}
\end{equation}
Here $t_{in}$, indicates the cosmological  times at the beginning
of inflation. It is clear that, in order to have a dark-radiation
dominated stage, we need to take $n=2$. On the other hand, note
that we have chosen to work with a power-law inflation. We could
relax this assumption, but the main consequences will not change.
Then, imposing $\ddot{a}>0$ leads to the constraint $p=C_{2}/2
>1$.

We assume that, either in the dark-radiation era as in the
inflationary era, the bulk-brane exchange energy term, $T$,
becomes low enough, so that, it does not affect the evolution of
the universe. For instance, in   the dark radiation period, we
have that $H \sim t^{-1}\sim a^{-2}$, and thus, the interaction
term falls as $T \sim a^{-(2+\alpha)}$ (see Eq. (\ref{interac})).
Comparing this term with that corresponding to energy density,
$a^{-3(1+\omega)}$, which starts to dominate when inflation comes
into play, and since we have taken $\alpha > 3(1+\omega)$, then,
we take the interaction term, $T$, to be negligible in this
period. Note that this situation agrees with that in which the
last term in Eq. (\ref{ansatz}) is associated to the solution of
the inhomogeneous part of the energy density field Equation, Eq.
(\ref{roeq2}). Thus, with these assumptions in mind, the onset of
high energy inflationary scenarios will be discussed in detail.

During the dark radiation dominated phase, $H^2 \simeq 2\gamma
\chi$, with $\chi$ dominated by the last term of
Eq.(\ref{chisol}),  and using Eq.(\ref{arad}), we get
\begin{equation} \label{Drel}
D \simeq \frac{a_{in}^{4}\,p^{2}\,}{2\gamma\,t_{in}^2}.
\end{equation}

Inflation starts at the high energy  in which $H^2 \simeq \beta
\rho^2$, where we have introduced the constant
$\beta\equiv\frac{\gamma}{\sigma}$. From this expression we get
that
\begin{equation}
\dot{H}\simeq\beta^{1/2}\,\dot{\rho}.\label{H}
\end{equation}
On the other hand, with the energy density given by the
homogeneous term of Eq. (\ref{ansatz}), $\rho \sim
\frac{B}{a^{3(1+\omega)}}$, we have
\begin{equation} \label{hereg}
a_{in} \simeq \left[3(1+\omega)\sqrt{\beta}B
\,t_{in}\right]^{1/3(1+\omega)}.
\end{equation}

 In the following, an homogenous single scalar field $\phi$ confined
 to the brane will be considered, whose energy density can lead to the accelerated
 expansion of the universe (inflation). Neglecting spatial gradients, the
 energy density and the pressure of the inflaton field are given by
\begin{equation}
\rho=\frac{\dot{\phi}^2}{2}+V(\phi)\,,\;\;\;\;\;\;P=\frac{\dot{\phi}^2}{2}-V(\phi),\label{rho}
\end{equation}
where $V(\phi)$ is the potential energy of the inflaton. The
evolution Equation of the inflaton field, $\phi$, is obtained
inserting Eqs. (\ref{rho}) into Eq. (\ref{1}), neglecting the
interaction term.

From Eqs. (\ref{1}), (\ref{H}) and (\ref{rho}) we obtain
\begin{equation}
\dot{\phi}^2=-\frac{\dot{H}}{3\,H\,\beta^{1/2}}=\frac{1}{3\,\beta^{1/2}\,t}.\label{df}
\end{equation}
Using this latter expression together with Eq.(\ref{rho}), we
obtain that the exact potential energy of the scalar field $\phi$
for power-law inflation becomes
\begin{equation} \label{hepot}
V(\phi)=\frac{1}{\beta^{1/2}}\,\left[H+\frac{\dot{H}}{6\,H}\right]=
\frac{2\,(6\,p-1)}{9\beta}\,\frac{1}{(\phi-\phi_{in})^2},
\end{equation}
where $\phi_{in}$, indicates the value of the scalar field at the
beginning of inflation. Of course, the form of this potential has
to be considered just as an approximation of a more complex
potential for the interval $\phi_{in} \leq \phi \leq \phi_{f}$.

In the following, the subscripts  $*$ and $f$ are used to denote
to the epoch when the cosmological scales exit the horizon (as
previously specified) and the end of  inflation, respectively. The
scalar perturbations for our model will be studied. We introduce
comoving curvature perturbations,
$\cal{R}=\psi+H\delta\phi/\dot{\phi}$, where $\delta\phi$ is the
perturbation of the inflaton field, $\phi$.  The power spectrum
associated to curvature perturbations is given by \cite{Bar}

\begin{equation}
{\cal{P}_R}\simeq\left.\frac{H^2}{\dot{\phi}^2}\left(\frac{H}{2\pi}\right)^2\right|_{k=k_*}\simeq
\frac{3\,\beta^{1/2}\,p^4}{4\,\pi^2\,t^3_{*}},\label{dp}
\end{equation}
where we have used Eq.(\ref{df}). Here, $k_*$  is referred to
$k=Ha$, the value when the universe scale  crosses the Hubble
horizon during inflation. The recent WMAP five-year results
\cite{wmap5} give the value for the scalar curvature spectrum
$P_{\cal R}(k_*)\simeq 2.4\times\,10^{-9}$ with $k_{*} = 0.002$
Mpc$^{-1}$.

On the other hand, the generation of tensor perturbations during
inflation would produce  gravitational waves and these
perturbations in cosmology are more involved, since gravitons
propagate in the bulk. The amplitude for tensor perturbations was
evaluated in Ref.\cite{t}
\begin{equation}
{\cal{P}}_g=48\gamma\,\left(\frac{H}{2\pi}\right)^2\;F^2(x)
\simeq\frac{12\gamma}{\pi^2}\,\frac{p^2}{t^2}\,F^2(x),\label{ag}
\end{equation}
where $x=Hm_p\sqrt{3/(4\pi\lambda)}$ and
$$
F(x)=\left[\sqrt{1+x^2}-x^2\sinh^{-1}(1/x)\right]^{-1/2}.
$$
Here, the function $F(x)$ appears from the normalization of a
zero-mode. The spectral index $n_g$ is given by $
n_g=\frac{d{\cal{P}}_g}{d\,\ln
k}=-\frac{2x_{,\,\phi}}{N_{,\,\phi}\,x}\frac{F^2}{\sqrt{1+x^2}}$.

From expressions (\ref{dp}) and (\ref{ag}) we may write  the
tensor-scalar ratio as
\begin{equation}
R=\left.\left(\frac{{\cal{P}}_g}{P_{\cal R}}\right)\right|_{k=k_*}
\simeq \frac{16\,\gamma\,t_*}{\beta^{1/2}\,p^2}\,F^2(t_*).
\label{Rk}\end{equation}

Combining  WMAP five-year data\cite{astro} with the Sloan Digital
Sky Survey  (SDSS) large scale structure surveys \cite{Teg},  an
upper bound for $R$ is found given by $R(k_*\simeq$ 0.002
Mpc$^{-1}$)$ <0.28\, (95\% CL)$, where $k_*\simeq$0.002 Mpc$^{-1}$
corresponds to $l=\tau_0 k\simeq 30$,  with the distance to the
decoupling surface $\tau_0$= 14,400 Mpc. The SDSS  measures galaxy
distributions at red-shifts $a\sim 0.1$ and probes $k$ in the
range 0.016 $h$ Mpc$^{-1}$$<k<$0.011 $h$ Mpc$^{-1}$. The recent
WMAP five-year results give the values for the scalar curvature
spectrum $P_{\cal R}(k_*)\simeq 2.4\times\,10^{-9}$ and the
scalar-tensor ratio $R(k_*)=0.055$. These values we will be used
to set constrains on the parameters of our model.

During inflation, from Eq. (\ref{ansatz}) we know that $B\simeq
\rho_{*}a_{*}^{3(1+\omega)}$ at the crossing time and from
Eq.(\ref{power}) we find that
\begin{equation}\label{rel1}
B \simeq
\frac{k_*^{3(1+\omega)}}{\beta^{1/2}}\,\left(\frac{t_*}{p}\right)^{2+3\omega}.
\end{equation}
So, having a value for $t_*$, we get an estimation for $B$, and
through Eqs. (\ref{Drel}) and (\ref{hereg}) we obtain a value for
the parameter $D$. The only constant that remains undetermined is
$a_{in}$, however using that $k_* = a_* H_*$, we can write
\begin{equation}\label{rel2}
a_{in} \simeq \,\frac{k_*\,t_{in}^p\; t_{*}^{1-p}}{p}.
\end{equation}
From Eq. (\ref{dp}), and in the case of power-law inflation  we
get
\begin{equation}\label{rel3}
t_* \simeq \left(\frac{3\sqrt{\beta}\,p^{4}}{4 \pi^2 {\cal{P}_R}}
\right)^{\frac{1}{3}}.
\end{equation}
The WMAP five-year data favors the tensor-scalar ratio $R\simeq
0.055$  and  from Eqs. (\ref{Rk}) and (\ref{rel3}) we obtain a
relation between the  parameters $\beta$ and $p$. In particular,
for $p=2$ we get the value for the brane tension
$\beta\simeq0.52\times 10^{10}m_p^{-6} $. For the case in which
$p=10$, we obtain $\beta\simeq 0.15\times 10^{3}m_p^{-6} $. Here,
we have used  the WMAP five year data where $P_{\cal R}(k_*)\simeq
2.4\times 10^{-9}$ and $k_*\simeq$0.002 Mpc$^{-1}$.

Finally, from Eqs.(\ref{Drel}), (\ref{rel2})  and (\ref{rel3}), we
obtain an estimation for the parameter $D$ associated to the dark
radiation energy density given by
\begin{equation}
D=\frac{k_*^4\;p^{\frac{2(5-8p)}{3}}}{2\,\gamma}\,
\left[\frac{3\sqrt{\beta}}{4\pi^2\,{\cal{P}_R}}\right]^{\frac{4(1-p)}{3}}\,t_{in}^{4p-2}.\label{d}
\end{equation}

\begin{figure}[th]
\includegraphics[width=3.0in,angle=0,clip=true]{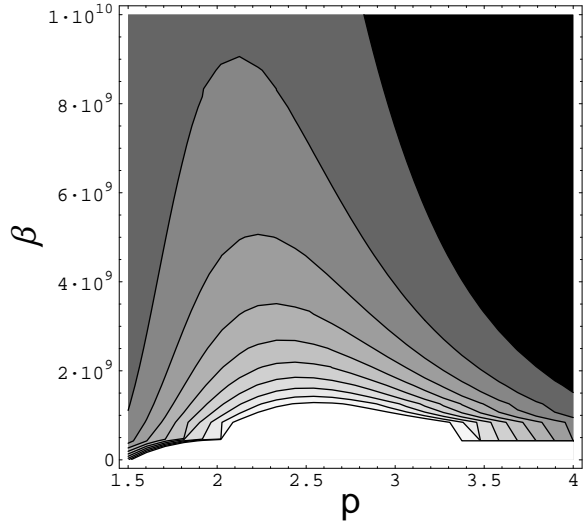}
\caption{ Contour plot for the dimensionless values of the
$D/k_*^4$  as a function of the parameters $\beta$ and $p$, fitted
from the cosmological times $t_{in}=10^5\,/m_p$. \label{cou2}}
\end{figure}

In  Fig.\ref{cou2} contours curves corresponding to the same
dimensionless number  $D/k_*^4$ are plotted, as well as
different combinations of the $\beta$ and $p$ parameters according
to Eq.(\ref{d}). Here, we have taken $t_{in}=10^5/m_p$. From this
plot, we see that, we can obtain the value of $D/k_*^4$, for a
given values of $\beta$ and $p$ parameters. In particular, for
$p=2$ and $\beta=0.52\times 10^{10} m_p^{-6}$ we get
$D/k_*^4=4.46\times 10^{10}$. For $p=10$ and $\beta=0.15\times
10^{3} m_p^{-6}$ we have $D/k_*^4=1.03\times 10^{36}$.

\section{Summary}

A  possible transition from a dark radiation period to an
inflationary phase has been here studied, based on a
Randall-Sundrum II type brane-world. During the dark radiation
period the energy density becomes $\chi \sim a^{-4}$, while in the
inflationary phase the energy density associated to the inflaton
field take the form $\rho\sim a^{-3(1+\omega)}$. During these
periods the scale factor $a$ evolves as $t^{1/2}$ and $t^p$ (with
$p>1$) in the dark radiation period and the inflationary phase,
respectively. The explicitly expression that connects these two
patches is $a(t)= a_{in} \left[ \sinh(C_{2}\ln
(1+t/t_{in}))\right]^{1/n}$, where $t_{in}$ represent the time
when inflation begins. It is considered that the bulk-brane energy
exchange, $T(\rho)$, plays no role in these two periods.

During the inflationary era the scale factor goes like
$a(t)=a_{in}\,(t/t_{in})^{p}$ with the parameter $p$ satisfying
$p>1$ in order to have $\ddot{a}>1$. This period is quite
interesting since we could fix some parameters appearing in the
model by using astronomical data  (coming, for instant, from WMAP
5 year). Actually, we succeeded in finding a relationship among
the parameters related to the dark radiation energy ($D$), the
brane tension parameter ($\beta$) and the power law parameter
($p$) for a given value of $t_{in}$. The relation of this
parameters becomes given by the Eq. (\ref{d}), i.e.
$D=\frac{k_*^4\;p^{\frac{2(5-8p)}{3}}}{2\,\gamma}\,
\left[\frac{3\sqrt{\beta}}{4\pi^2\,{\cal{P}_R}}\right]^{\frac{4(1-p)}{3}}\,t_{in}^{4p-2}$.
This relation is summarized in the plot showed in Fig.\ref{cou2}.
The interesting point here is that if we know the brane tension
parameters, and $p$ from Eqs.(\ref{Rk}) and (\ref{rel3}), we could
fix an appropriated value for the parameter $D$ associated to the
dark radiation energy density. This might seem to be a very
particular result, since the assumption of power law inflation is
by itself a particular perform, but we think that relaxing this
assumption our main results will not change very much. We hope to
approach this problem in the near future.


\section*{Acknowledgments}

S.d.C. was supported from Comisi\'{o}n Nacional de Ciencias y
Tecnolog\'{\i}a (Chile) through FONDECYT Grants 1070306 and
1080530 and by the PUCV-DGI/123.787. R.H. was supported by the
Programa Bicentenario de Ciencia y Tecnolog\'{\i}a" through the
Grant  Inserci\'{o}n de Investigadores Postdoctorales en la
Academia" N0 PSD/06.



\begin{references}
\bibitem{g4} The case of low-energy regime, i.e., $\rho < \sigma$,
was studied in Ref.\cite{g3}.
\bibitem{20} Here, at difference of Ref.\cite{vh}, the solution
(20) is not a solution of the complete gravity-dark
radiation-inflation system, rather it is a simple interpolation
between the dark radiation and inflationary periods.
\end{references}
\end{document}